\documentclass[11pt]{iopart}
\usepackage{iopams}
\usepackage{graphicx}
\usepackage{textcomp}
\usepackage[utf8]{inputenc}
\usepackage{setspace}
\newcommand*{\newblock}{}
\renewcommand{\vec}{\bi}

\begin{document}

\title[Slow short-time phospholipid motions, multiple scattering, and fitting artifacts]{The slow short-time motions of phospholipid molecules with a focus on the influence of multiple scattering and fitting artifacts}

\author{Sebastian Busch$^1$ and Tobias Unruh$^{1,2}$}

\address{$^1$Technische Universit\"at M\"unchen, Physik Department E13 and Forschungs-Neutronenquelle Heinz Maier-Leibnitz (FRM~II), Lichtenbergstr.~1, 85748~Garching bei M\"unchen, Germany}
\ead{sbusch@ph.tum.de}

\address{$^2$Universit\"at Erlangen, Lehrstuhl f\"ur Kristallographie und Strukturphysik, Staudtstr.~3, 91058~Erlangen, Germany}
\ead{tobias.unruh@krist.uni-erlangen.de}

\begin{abstract}
Quasielastic neutron scattering is a powerful tool for the study of non-periodic motions in condensed matter as a detailed line shape analysis can give information about the geometry and rate of the scatterers' displacements. Unfortunately, there are also a number of artifacts which can masquerade as signatures of motions and can therefore lead to erroneous results. Their influence on the evaluation of the motions of the phospholipid dimyristoylphosphatidylcholine (DMPC) is discussed. On a 60\,ps time scale, the long-range motion of the molecules has a flow-like character with similar velocities above and below the main phase transition. It is proposed that the concepts of dynamical heterogeneities and ``floppy modes'' developed in glass physics provide a framework to explain the observed behaviour.
\end{abstract}

\submitto{\JPCM}
\maketitle

\twocolumn

\section{Introduction}

\begin{figure}
\includegraphics[width=\columnwidth]{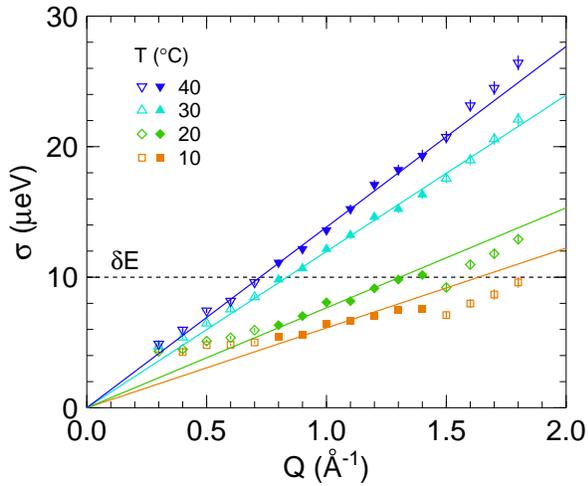}%
\caption{Width of the narrow Gaussian line of a three-component fit to the spectra of the fully hydrated phospholipid dimyristoylphosphatidylcholine (DMPC), very similar to the results published before~\cite{Busch10flow} where it was speculated that the levelling off of the widths at low temperatures and small $Q$ could either be a signature of confinement of the molecules or an artifact. The distance between two data points in the spectra, $\delta E = 10$\,\textmu{}eV, is indicated in the figure. The energy resolution is about 60\,\textmu{}eV, corresponding to an observation time of about 60\,ps. The solid lines are fits to the filled points in the medium-$Q$ region.}
\label{fig:fit_w1}
\end{figure}

Recently, a quasielastic neutron scattering (QENS) experiment on the fully hydrated phospholipid dimyristoylphosphatidylcholine (DMPC) was published~\cite{Busch10flow}. It corroborated collective flow-like motions as a novel view of the long-range motion of phospholipid molecules that had been put forward by molecular dynamics simulations~\cite{Falck08}. A fit to the data showed that the component which was assigned to the long-range flow-like motion of the molecules levelled off at temperatures below the main phase transition (24\textcelsius{}) and low values of $Q$ as shown in \fref{fig:fit_w1}. This was discussed to be a sign of a confinement of the molecules during the 60\,ps observation time in a cage of their neighbours -- or an artifact~\cite{Busch10flow,Swenson02,Zorn10}.

The macroscopic diffusion of phospholipid molecules is known to decrease by about one order of magnitude when the membrane is cooled below the main phase transition~\cite{Ghosh10}. In contrast, the high-$Q$ part of the broadening caused by the long-range motion shown in \fref{fig:fit_w1} decreases much less. The confinement length that was extracted from the position of the deviation of the linear behaviour at low $Q$ to about 9\,\AA{} coincides quite well with the molecular repeating distance in the membrane. This made a flow-like motion of the single molecules until they hit the cage of the neighbouring molecules seem reasonable.

However, it was speculated already there that this might be an artifact as the broadening is very small compared to the instrumental resolution. A caging effect should also not only cause a plateau of the quasielastic broadening but simultaneously an emerging elastic $\delta(\omega)$ component which was not observed. On a physical level, it is not clear why the flow-like, collective motions of the molecules together with their caging neighbours above the phase transition should persist at low temperatures for the single molecules within the cages.

The present contribution aims to clarify the picture of the motions of phospholipid molecules on a 60\,ps time scale below the main phase transition. The two likely artifacts that could simulate the observed plateau of the broadening at low $Q$ are studied: multiple scattering and fit artifacts.

Only two years after the collection of the first neutron diffraction patterns ever in 1946~\cite{Shull95}, multiple scattering was identified as the source of a spurious background contribution~\cite{Wollan48}. A few years later, efforts to treat multiple scattering (semi-)analytically were started~\cite{Vineyard54,Skoeld72,Sears75,Zorn02} and later complemented by Monte Carlo simulations~\cite{Copley74,Wuttke93,Wuttke96,Russina99,Mezei99,Wuttke00mss,Trouw03,Hugouvieux04}.

However, still by far most quasielastic neutron scattering experiments study samples with very high transmission in order to be able to neglect rather than correct multiple scattering. The transmission needed to justify this neglect is not agreed on in the literature -- some authors feel safe with samples with a transmission of 99\%~\cite{Pieper09}, most adjust their samples to achieve a transmission of about 90\%~\cite{Stadler08water,Trapp10} but others still feel comfortable with samples with a transmission as low as 70\%~\cite{Thompson06}.

It is also often unclear what to expect from uncorrected multiple scattering. An artificial broadening due to multiple scattering was for example assumed to yield a diffusion coefficient that was higher~\cite{Aliotta02} than the generally accepted value~\cite{Spehr10phd,Spehr11}. Others suggested that this broadening results in a $Q$ dependence of the line width as shown in \fref{fig:fit_w1} which is normally interpreted as confined diffusion~\cite{Zorn10}. Offsets of the line width and the elastic incoherent structure factor (EISF) are also often observed~\cite{Zorn02,Gaspar08,Stadler08temp,Busch10flow} some of which were shown to be due to multiple scattering.

Whereas the neutron scattering community is well aware of the problem of multiple scattering,
fitting artifacts are widely ignored which can lead to erroneous results: one crucial step is the numerical convolution of the theoretical fit function with the measured instrumental resolution~\cite{Doster10}. The instrumental resolution gets more important as neutron spectroscopy is employed for the study of very slow motions because the line broadening caused by the sample gets small.

It is even sometimes said that the instrumental resolution should be seen as a lower limit for the detectable broadening. However, while this is a valid rule of thumb for localised motions, many measurements of long-range motions which cause a broadening of the whole line are done below this limit: from neutron scattering studies of diffusion in water~\cite{Teixeira85} over proteins~\cite{Stadler08temp} to metallic liquids~\cite{Meyer98} or X-ray scattering studies of glass formers~\cite{Sette98}.

In the following, a short repetition of the basic assumptions of neutron scattering data evaluation is presented, highlighting the points where multiple scattering and fit artifacts come into play. After a description of the studied systems, the effects of multiple scattering and the point density used for the fit are described generally enough to be useful also for other experiments. The consequences for the evaluation of long-range and localized motions are presented and with this knowledge, a coherent picture for the phospholipid dynamics is proposed.
\section{Concepts of QENS data evaluation}

The evaluation of QENS data relies vitally on the fact that the recorded number of neutrons as a function of solid angle and energy transfer, the double differential cross section, can be transformed into the scattering functions $S^\mathrm{exp}_\mathrm{coh,inc}(\vec{Q},\omega)$ via~\cite{vanHove54,Furrer09,Squires96}
\begin{eqnarray}
    \fl
    \left( \frac{\rmd^2 \sigma}{\rmd \omega \rmd \Omega} \right) = \frac{N \sigma_\mathrm{coh}}{4 \pi} \frac{k_f}{k_i} S_\mathrm{coh}^\mathrm{exp}(\vec{Q},\omega) + \nonumber\\
    \fl
    \qquad\qquad
    \frac{N \sigma_\mathrm{inc}}{4 \pi} \frac{k_f}{k_i} S_\mathrm{inc}^\mathrm{exp}(\vec{Q},\omega) \quad .
    \label{equ:ddcs}
\end{eqnarray}
In this relation, it is assumed that a monochromatic beam of neutrons was scattered at most once in the sample and that the scattering lengths of the atoms do not correlate with their positions. It is often further simplified by neglecting the coherent scattering contributions of hydrogen-rich samples and using only the absolute value of the scattering vector, $Q$, for powder samples.

To account for the non-perfect mono\-chroma\-ticity of the incident neutron beam and the finite energy resolution of the spectrometer, the experimentally obtained scattering function $S_\mathrm{inc}^\mathrm{exp}(Q,\omega)$ is understood as the convolution of the instrumental resolution function $R(Q,\omega)$ with the theoretical scattering function $S_\mathrm{inc}^\mathrm{theo}(Q,\omega)$:
\begin{equation}
    \fl
    S_\mathrm{inc}^\mathrm{exp}(Q,\omega) = S_\mathrm{inc}^\mathrm{theo}(Q,\omega) \otimes R(Q,\omega) \quad .
    \label{equ:SconvR}
\end{equation}
For the analysis of QENS data, this convolution is carried out only in the energy-direction. As the real resolution of the spectrometer does not strictly follow an analytical function, the integral form of this convolution
\begin{eqnarray}
    \fl
    S_\mathrm{inc}^\mathrm{exp}(Q,\omega) = \int_{-\infty}^{\infty} S_\mathrm{inc}^\mathrm{theo}(Q,\omega') \cdot \nonumber\\
    \fl
    \qquad\qquad
    R(Q,\omega-\omega') \ \rmd \omega'
    \label{equ:convint}
\end{eqnarray}
is mostly replaced by the discrete equivalent
\begin{eqnarray}
    \fl
    S_\mathrm{inc}^\mathrm{exp}(Q,\omega_n) = \sum_{m=1}^N S_\mathrm{inc}^\mathrm{theo}(Q,\omega_m) \cdot \nonumber \\
    \fl
    \qquad\qquad
    R(Q,\omega_n-\omega_m) \cdot \delta \omega
    \label{equ:convsum}
\end{eqnarray}
where $N$ is the total number of points of a spectrum in the energy direction and $\delta \omega$ the width of a bin.

The relevant motions of the scatterers are then assumed to be independent of each other so that the scattering function can be approximated as a composition of e.\,g.\ translational, rotational, and vibrational contributions~\cite{Bee88},
\begin{eqnarray}
    \fl
    S_\mathrm{inc}^\mathrm{theo}(Q,\omega) = S_\mathrm{inc}^\mathrm{trans}(Q,\omega) \otimes \nonumber\\
    \fl
    \qquad\qquad
    S_\mathrm{inc}^\mathrm{rot}(Q,\omega) \otimes S_\mathrm{inc}^\mathrm{vib}(Q,\omega) \quad .
    \label{equ:SQwBee}
\end{eqnarray}

The amount of physics \emph{and approximations} contained in these equations is so overwhelming that it seems safe to postulate that by far most neutron scatterers (including the authors) settle for using these approximations after following the calculations a few times.

In this contribution, we discuss the influence of two of these steps: (a)~The assumption that all detected neutrons were scattered only once in the sample and (b)~the correction of the limited instrumental resolution by the discrete convolution of the theoretical scattering function with an experimentally obtained instrumental resolution function.

\section{Experimental}

\subsection{Materials}

A vanadium hollow cylinder standard with an outer radius 11.75\,mm and wall thickness 0.6\,mm was measured at 20\,\textcelsius{}. The transmission was calculated by numerical integration of the path length of the neutrons through the scatterer in transmission geometry and assuming an exponential decay of the neutron intensity in the medium~\cite{Frida1} to 93\%.

For all samples, hollow cylindrical aluminium sample containers~\cite{Wuttke99,Wuttke00sac} with an inner radius of the outer aluminium layer of 11.25\,mm and varying radii of the inner aluminium layer were used. The choice of the insert determined the gap size and therefore the sample layer thickness.

\begin{figure}
\includegraphics[width=\columnwidth]{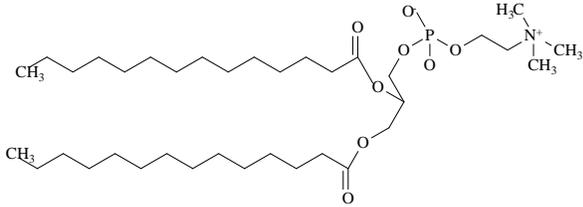}
\caption{Molecular structure of DMPC.}
\label{fig:dmpc}
\end{figure}

The phospholipid dimyristoylphosphatidylcholine (DMPC, cf.\ \fref{fig:dmpc}) was obtained from Lipoid GmbH, Ludwigshafen (Germany) and used as received. It was hydrated at 40\,\textcelsius{} via the gas phase with D$_2$O 99.90\% from Euriso-Top, Gif-sur-Yvette (France) for at least 48\,h until a clear phase formed. Subsequently, further D$_2$O was added up to 50 weight~\% to ensure full hydration during the whole experiment. With a gap size of 0.2\,mm, a calculated transmission of 85\% was obtained. The measurements were performed at 10\,\textcelsius{}, 20\,\textcelsius{}, 30\,\textcelsius{}, and 40\,\textcelsius{}. Also pure D$_2$O was measured at these temperatures.

Pure H$_2$O, already previously used for multiple scattering studies~\cite{Slaggie67,Slaggie69}, was taken from a Barnstead EASYpure II ultrapure water system with a resistivity of $\rho \geq 18.2$\,M$\Omega\cdot$cm and filled into the sample containers with gap sizes of 0.1\,mm, 0.2\,mm, 0.3\,mm, 0.5\,mm, 1.0\,mm, 2.0\,mm, and 3.0\,mm. These samples have calculated transmissions of 85\%, 72\%, 62\%, 46\%, 23\%, 6.4\%, and 2.2\%, respectively -- the mean free path length of a neutron in H$_2$O is about 2\,mm. The measurements were performed at 20\,\textcelsius{}.

\subsection{Measurements and data treatment}

All experiments were performed at the cold neutron time-of-flight spectrometer TOFTOF \cite{Unruh07tof,Unruh08tof} at the Forschungs-Neutronenquelle Heinz Maier-Leibnitz (FRM~II), Garching bei M\"{u}nchen, Germany. The spectrometer is equipped with detectors in an angular range from 7.5\textdegree{} to 140\textdegree{}.

An incident neutron wavelength of 6\,\AA{} was used, hereby giving for quasi-elastic scattering access to $Q$ values from 0.3\,\AA$^{-1}$ to 1.8\,\AA$^{-1}$. The maximal momentum transfer of elastically scattered neutrons at this wavelength is $\hbar \cdot$2.1\,\AA$^{-1}$ for backscattered neutrons. An instrumental resolution of about 60\,\textmu{}eV (full width at half maximum) was reached with a chopper rotation speed of 12000\,rpm which required a frame overlap ratio of 4. The corresponding observation time is about 60\,ps~\cite{Unruh08alk}.

Today's normal way of histogramming the data online during the experiment adds a restriction to the rebinning of the data.
The online histogramming was done into 1024 time channels, resulting in a width of about 7.5\,\textmu{}eV of the bins at zero energy transfer. These data points were then rebinned into bins with a constant step in energy of $\delta E$=10\,\textmu{}eV during data reduction into slices of constant $Q$. Additionally, the phospholipid was also measured with an online histogramming into 4096 time channels, resulting in a width of about 1.9\,\textmu{}eV at zero energy transfer and a constant step of $\delta E$=2\,\textmu{}eV after rebinning.

The data were normalized to the scattering of the vanadium standard to correct for different detector efficiencies. An empty can measurement was subtracted from all measurements and the D$_2$O spectra were subtracted pro-rata from the ones of the phospholipid at the corresponding temperatures. The obtained spectra were not normalised to the amount of sample in the beam or the total scattered intensity.

Diffraction patterns $\rmd \sigma / \rmd \Omega(2\theta)$ were obtained by integrating the double differential cross section over all energies. Spectra were obtained by regrouping the data into slices of constant $Q$~\cite{Frida1}. In energy, the spectra were evaluated in the region of $\Delta E = \hbar \omega$ from $-1$\,meV to $+1$\,meV where positive values for the energy transfer denote an energy gain of the neutron during the scattering process(es).

The fits of the theoretically calculated scattering functions $S_\mathrm{inc}^\mathrm{theo}(Q,\omega)$ to the measured ones $S_\mathrm{inc}^\mathrm{exp}(Q,\omega)$ were numerically convolved with the instrumental resolution $R(Q,\omega)$ as determined by the measurement of the vanadium hollow cylinder standard.

In order to enhance the visibility of the inelastic part of the spectrum, a representation as the imaginary part of the dynamic susceptibility is also shown, the transformation was done using
\begin{equation}
    \fl
    \chi''(Q,\omega) = S_\mathrm{inc}^\mathrm{exp}(Q,\omega) / n(\omega)
    \label{equ:dyn_susc}
\end{equation}
where the scattering functions are divided by the Bose occupation factor
\begin{equation}
    \fl
    n(\omega) = 1 / \left( \exp\left[\hbar \omega / (k_\mathrm{B} T)\right] - 1 \right) \quad .
    \label{equ:bose}
\end{equation}

Intermediate scattering functions $I_\mathrm{inc}^\mathrm{exp}(Q,t)$ were obtained by cosine Fourier transform of the energy spectra and subsequent division of the sample data by the instrumental resolution as determined by the vanadium measurement. Again, the data were not normalised to $I_\mathrm{inc}^\mathrm{exp}(Q,t=0)=1$. Instead, an overall scaling factor $a$ was included into the fitting function.

\subsection{Fitting models}

The data of the phospholipid were fitted with a convolution of a Gaussian for the long-range flow-like motion with two localized diffusive contributions, each of the form
\begin{eqnarray}
    \fl
    S_\mathrm{inc}^\mathrm{theo}(Q,\omega) = A_0(Q) \cdot \delta(\omega) + \nonumber \\
    \fl
    \qquad
    (1-A_0(Q)) \cdot L(\Gamma_\mathrm{broad}(Q),\omega)
    \label{equ:localized}
\end{eqnarray}
where $L$ denotes a Lorentzian
\begin{equation}
    \fl
    L(\Gamma(Q),\omega) = \frac{\Gamma(Q)/\pi}{\Gamma(Q)^2+\omega^2} \quad .
    \label{equ:Lor}
\end{equation}
The different localized contributions were assigned to the motions of the atoms in the head and tail groups, respectively~\cite{Busch10flow}. The Gaussian then captures only the long-range motion of the whole molecule.
The Gaussian line shape with a line width of
\begin{equation}
    \fl
    \sigma(Q) = v \cdot Q
    \label{equ:flow_sig}
\end{equation}
results from the assumption of flow motions with Maxwell-Boltzmann distributed flow velocities around the most likely velocity $v$~\cite{Busch10flow}.

The spectra of H$_2$O were fitted with a common function in quasielastic neutron scattering, the sum of two Lorentzians which is motivated~\cite{Unruh08alk} by the convolution of a long-range diffusive motion
\begin{equation}
    \fl
    S_\mathrm{inc}^\mathrm{theo}(Q,\omega) = L(\Gamma_\mathrm{narrow}(Q),\omega) \quad ,
    \label{equ:longrange}
\end{equation}
with a localized motion \eref{equ:localized}. With a scaling prefactor $a$ accounting for the scattering power of the sample, this yields
\begin{eqnarray}
    \fl
    S_\mathrm{inc}^\mathrm{theo}(Q,\omega) = a \cdot L(\Gamma_\mathrm{narrow}(Q),\omega) \otimes \nonumber\\
    \fl
    \qquad\qquad
    \big[ A_0(Q) \cdot \delta(\omega) + \nonumber\\
    \fl
    \qquad\qquad
    (1-A_0(Q)) \cdot L(\Gamma_\mathrm{broad}(Q),\omega) \big] \nonumber\\
    \fl
    \qquad
    = a \cdot A_0(Q) \cdot L(\Gamma(Q)_\mathrm{narrow},\omega) + \nonumber\\
    \fl
    \qquad\qquad
    a \cdot (1-A_0(Q)) \cdot L(\Gamma(Q)_\mathrm{narrow} + \nonumber\\
    \fl
    \qquad\qquad
    \Gamma(Q)_\mathrm{broad},\omega)) \quad .
    \label{equ:SQw_2Lor}
\end{eqnarray}
A specific model for the localised motion, giving an expression for $A_0(Q)$ as used in previous works~\cite{Teixeira85} was avoided because of the distortion by multiple scattering~\cite{Bee88,Wuttke00mss,Busch10flow}.

For the evaluation of the intermediate scattering function, another fit function will be used which is also very common for the evaluation of QENS data: a stretched exponential,
\begin{equation}
    \fl
    I_\mathrm{inc}^\mathrm{theo}(Q,t) = a \cdot \exp \left[ -\left(t/\tau\right)^\beta \right] \quad .
    \label{equ:IQt_str}
\end{equation}
The alternative, a sum of two exponential decays, gives very similar results to the ones obtained from the fit of the scattering function with two Lorentzians as one would expect~\cite{Smuda08alk,Smuda09phd}. The resulting relaxation time is shown as mean relaxation time $\langle \tau \rangle = \tau \cdot \Gamma(\beta^{-1}) \cdot \beta^{-1}$ where $\Gamma$ is the Gamma function, reducing the dependence of $\tau$ on $\beta$~\cite{Wuttke00mss,Doster10}.

\section{Results}

\subsection{Multiple scattering}

This section is organized as follows: After looking qualitatively at the data and their distortion caused by multiple scattering, the two most common ways for the quantitative evaluation of QENS data will be presented: fits of a sum of Lorentzians to the scattering function $S_\mathrm{inc}^\mathrm{exp}(Q,\omega)$ and the fit of a stretched exponential to the intermediate scattering function $I_\mathrm{inc}^\mathrm{exp}(Q,t)$.

\subsubsection{The angular dependence of the scattered intensity.}

\begin{figure*}
\includegraphics[height=0.25\textwidth,width=0.3\textwidth,keepaspectratio=true]{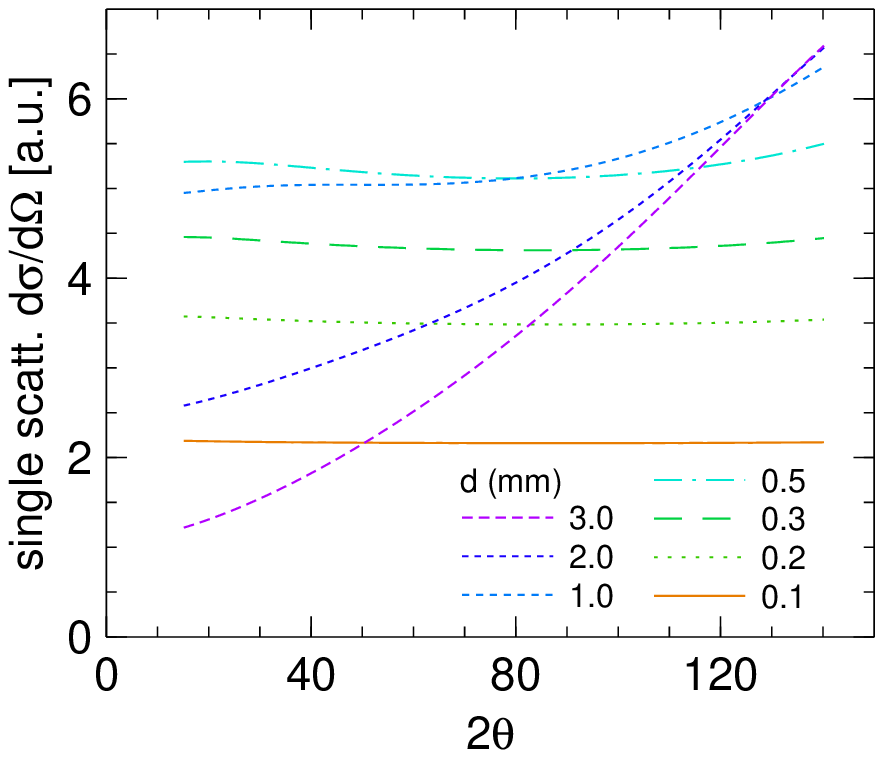}%
\hfill%
\includegraphics[height=0.25\textwidth,width=0.3\textwidth,keepaspectratio=true]{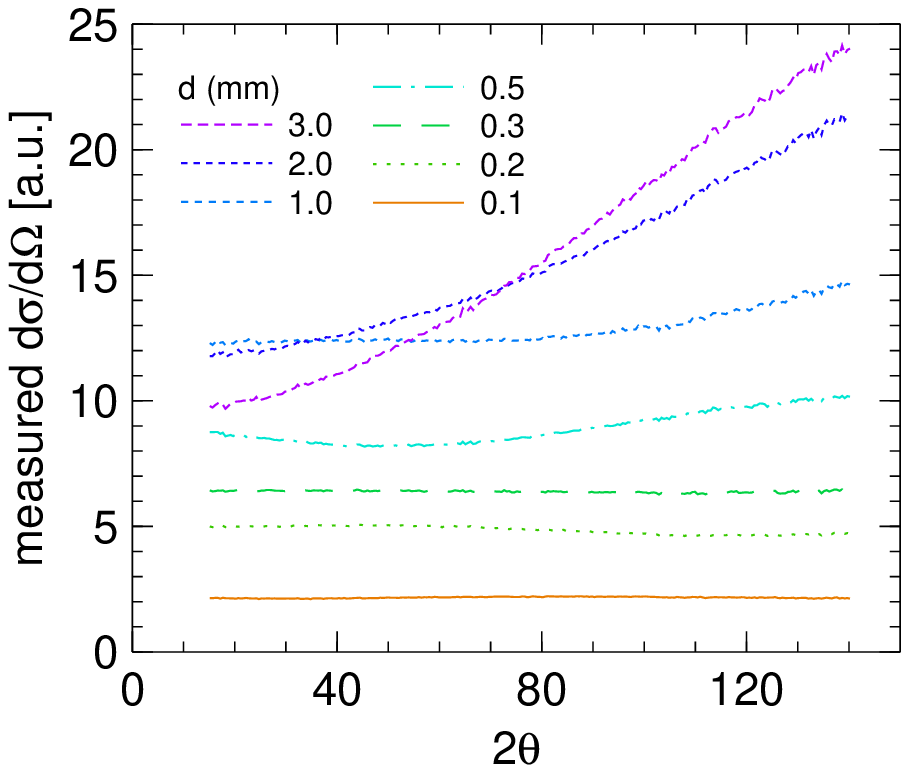}%
\hfill%
\includegraphics[height=0.25\textwidth,width=0.3\textwidth,keepaspectratio=true]{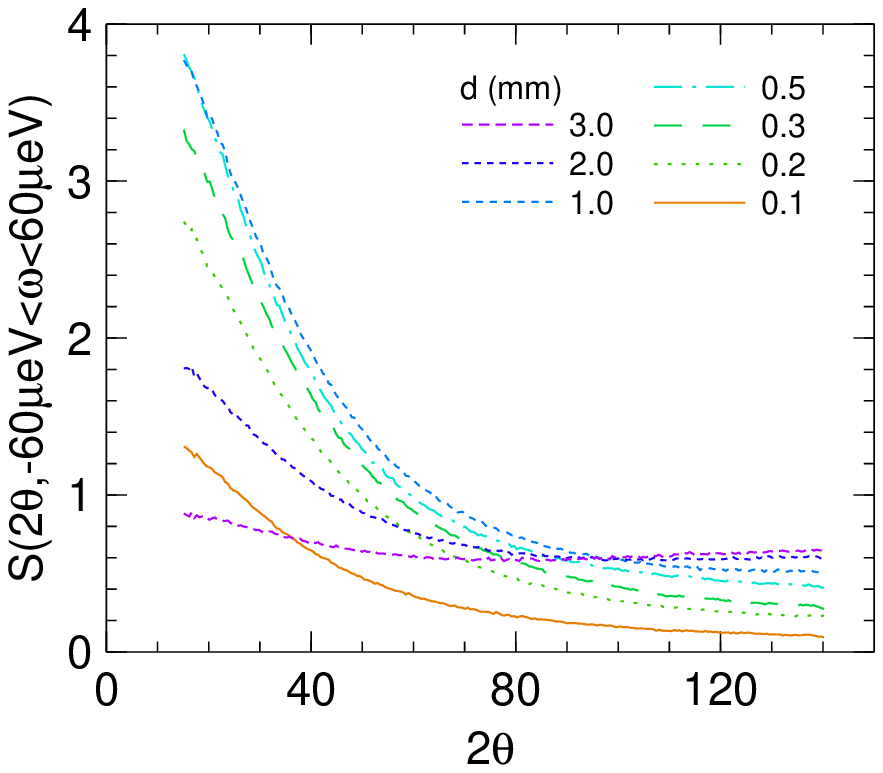}
\caption{From left to right: (a)~Calculated angular dependence of the differential cross section of H$_2$O, caused only by single-scattered neutrons. (b)~Measured angular dependence of the differential cross section of H$_2$O. (c)~Measured angular dependence of the elastic part of the scattering function of H$_2$O.}
\label{fig:ms_diff}
\end{figure*}

The algorithm of Paalman and Pings~\cite{Paalman62} was used to calculate the path lengths of the neutrons through the samples as a function of the scattering angle. Using the average scattering cross section, the probabilities for them to be scattered was determined from this length~\cite{Frida1}. Only those neutrons which were scattered exactly once were considered to be detected.
To facilitate comparability with the measured data, the resulting differential scattering cross sections were scaled with a factor that was chosen such that the calculated integral scattering cross section in the covered angular range is equal to the measured one for the 0.1\,mm sample, cf.~\fref{fig:ms_diff}a.

The energy-integrated scattering cross section, the diffraction pattern one would measure at a standard diffraction instrument, is shown in \fref{fig:ms_diff}b. Several differences to the calculated one are visible: (a)~The scattered amount of neutrons is much higher than expected from the calculation because double scattered neutrons are actually not lost. (b)~Increasing the thickness of the water layer generally also increases the scattered intensity. This increase is however very small for the thick samples -- the 3.0\,mm sample scatters hardly more, at small scattering angles even less, than the 2.0\,mm sample. (c)~The angular variation of the scattered intensity is lower than expected from the calculation.

The elastically scattered intensity determined by integration of an energy range of $\pm60$\,\textmu{}eV is shown in \fref{fig:ms_diff}c. It can easily be seen that the amount of elastically scattered neutrons does not scale linearly with the amount of sample -- the thickest sample even scatters less neutrons elastically than the thinnest at low angles.

\subsubsection{Intensity redistribution in $(Q,\omega)$-space.}

\begin{figure*}
\includegraphics[height=0.25\textwidth,width=0.3\textwidth,keepaspectratio=true]{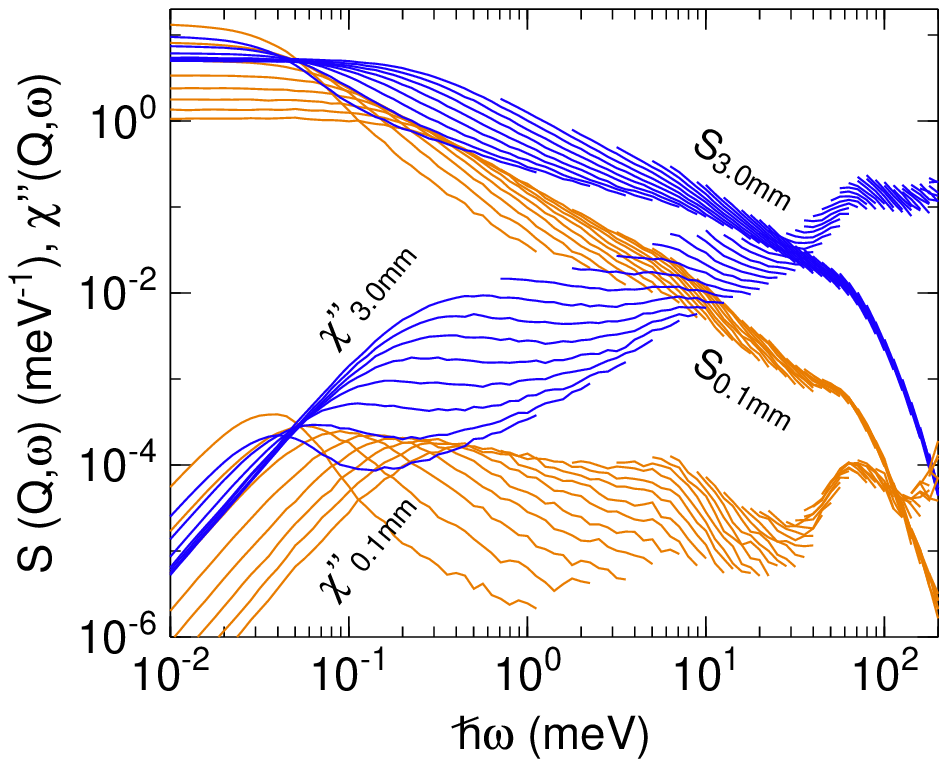}%
\hfill%
\includegraphics[height=0.25\textwidth,width=0.3\textwidth,keepaspectratio=true]{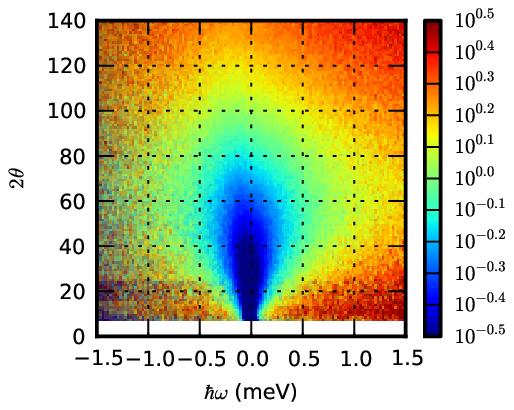}%
\hfill%
\includegraphics[height=0.25\textwidth,width=0.3\textwidth,keepaspectratio=true]{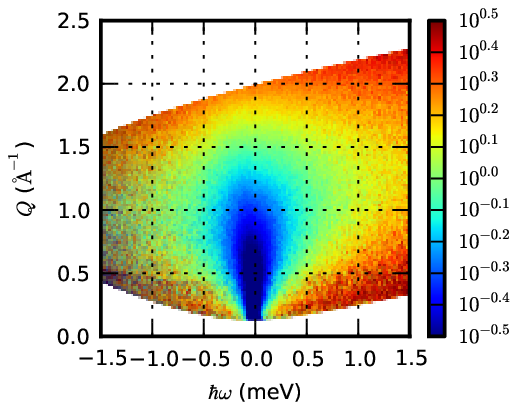}%
\caption{From left to right: (a)~The scattering function $S_\mathrm{inc}^\mathrm{exp}(Q,\omega)$ as well as the imaginary part of the dynamical susceptibility $\chi''(Q,\omega)$ for two H$_2$O samples (0.1\,mm and 3.0\,mm sample layer thickness) and values of $Q$ from 0.4\,\AA$^{-1}$ to 10.0\,\AA$^{-1}$ in steps of 0.2\,\AA$^{-1}$. Due to the dynamical range of the spectrometer, the data at low $Q$ do not reach high energy transfers and vice versa. (b)~The ratio of the normalized double differential cross sections of the 3.0\,mm measurement and the 0.1\,mm measurement. (c)~Shows the same as (b) after conversion from scattering angle $2\theta$ to $Q$.}
\label{fig:ms_inel}
\end{figure*}

Comparing the scattering function $S_\mathrm{inc}^\mathrm{exp}(Q,\omega)$ of the thinnest and the thickest sample in \fref{fig:ms_inel}a, it becomes clear that the elastic signal of the 3.0\,mm measurement depends much less on $Q$ than the one of the 0.1\,mm measurement, in agreement with \fref{fig:ms_diff}c. At higher energy transfers, the 3.0\,mm sample has up to one order of magnitude more signal than the 0.1\,mm measurement. This artificial gain in intensity distorts the dynamical susceptibility even more than the scattering function.

To look closer at the redistribution of the scattered intensity in the quasielastic region, the double differential scattering cross section of the 3.0\,mm measurement and the one of the 0.1\,mm measurement were independently normalized to the same total scattering cross section in the dynamical range. The data of the 3.0\,mm measurement were then divided by the ones of the 0.1\,mm measurement. The result is shown as a function of energy transfer and scattering angle in \fref{fig:ms_inel}b as well as a function of energy transfer and $Q$ in \fref{fig:ms_inel}c. Values smaller than $10^0$ designate areas where multiple scattering caused a loss of intensity, values bigger than $10^0$ those where additional, multiply scattered neutrons were detected.

The pattern shows a slight asymmetry with respect to the energy transfer. The lower intensity observed on the neutron energy loss side was attributed to the higher absorption probability for slower neutrons.

At low scattering angles, the signal of the 0.1\,mm measurement decreases more rapidly as a function of energy transfer than the signal of the 3.0\,mm measurement which is more smeared out, cf.\ \fref{fig:ms_inel}a. This leads to the high relative intensity at low scattering angles and large energy transfers. With increasing scattering angle, the shapes of the scattering functions become more similar to each other.

\subsubsection{Evaluation of the scattering function with two Lorentzians.}

\begin{figure*}
\includegraphics[height=0.2\textwidth,width=0.25\textwidth,keepaspectratio=true]{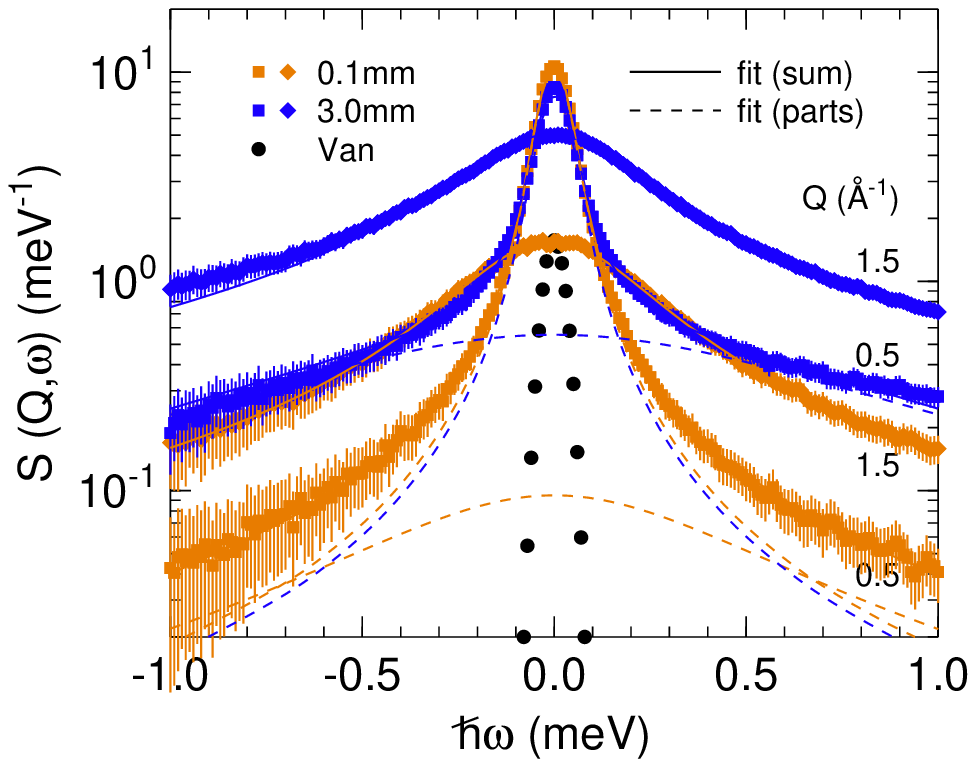}%
\hfill%
\includegraphics[height=0.2\textwidth,width=0.25\textwidth,keepaspectratio=true]{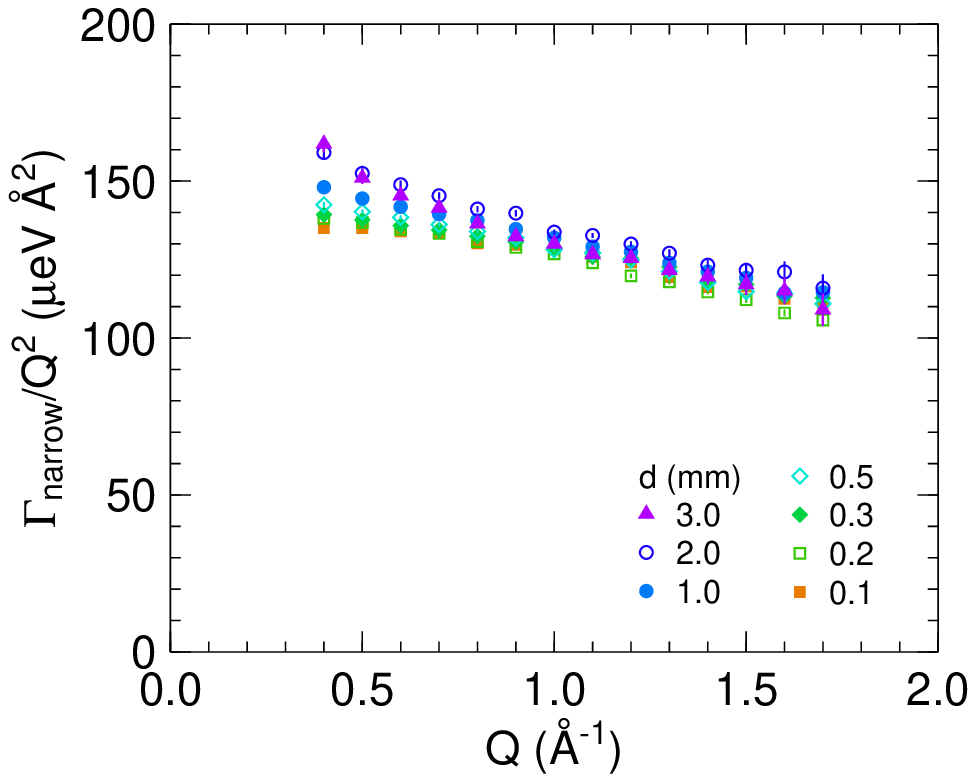}%
\hfill%
\includegraphics[height=0.2\textwidth,width=0.25\textwidth,keepaspectratio=true]{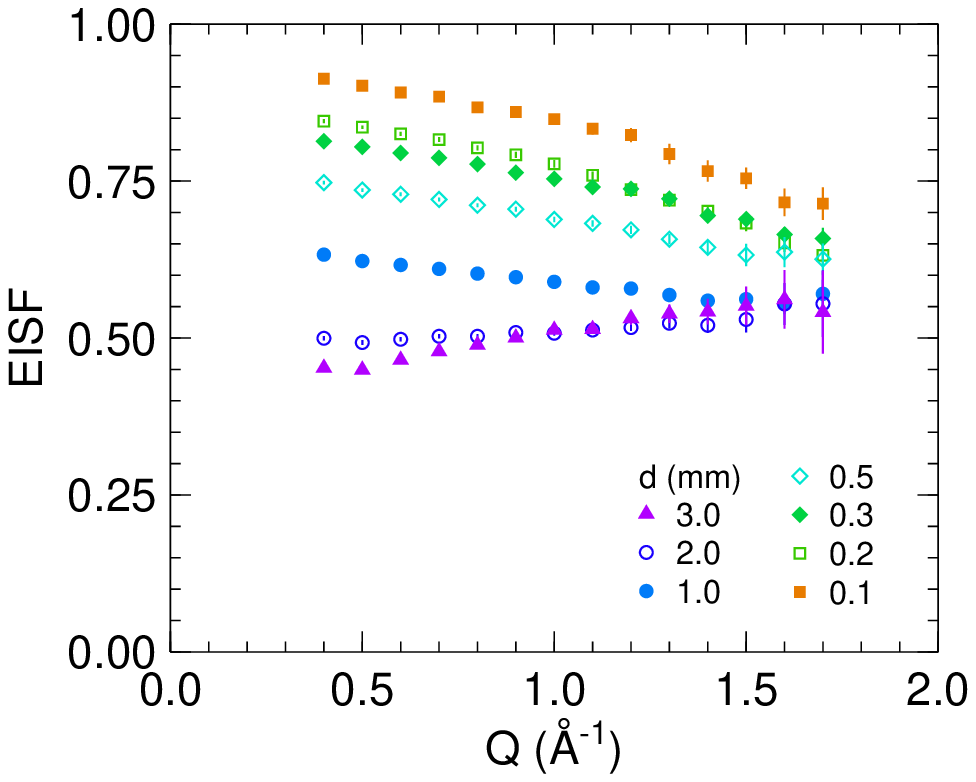}%
\hfill%
\includegraphics[height=0.2\textwidth,width=0.25\textwidth,keepaspectratio=true]{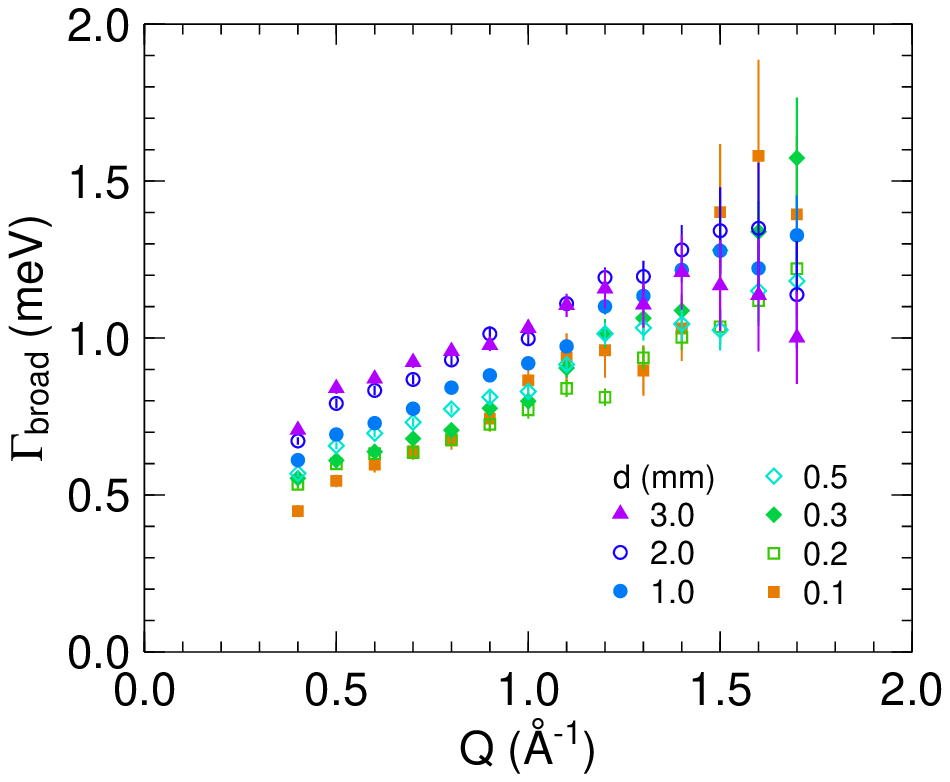}
\caption{From left to right: (a)~Scattering function of H$_2$O, shown at two $Q$ values and two sample layer thicknesses. Also shown is a vanadium measurement as a measure of the instrumental resolution and fits of the sum of two Lorentzians to the data (drawn-out lines). For the two measurements at low $Q$, also the two components of the fit are shown (dashed lines). (b)~The line width of the narrow of the two Lorentzians. (c)~The elastic incoherent structure factor $A_0$. (d)~The line width of the broad of the two Lorentzians.}
\label{fig:ms_SQw}
\end{figure*}

In the fit with two Lorentzians~\eref{equ:SQw_2Lor}, the properties of the narrow and the broad one, corresponding to long-range and localized motions, respectively, are discussed in the following.

The arguably most important observation in the presented measurements is shown in \fref{fig:ms_SQw}a and \fref{fig:ms_SQw}b: the line width of the narrow Lorentzian is largely unaffected by multiple scattering. This means that a reliable determination of the diffusion coefficient is possible even when a massive amount of multiple scattering is present. This will be detailed in the discussion.

The intensity distribution between the two Lorentzians, the EISF, follows for the thin samples quite well the expected Gaussian-like decrease with $Q$, see \fref{fig:ms_SQw}c. However, it is already visible in the 0.1\,mm measurement that the EISF does not extrapolate to 1 as $Q$ approaches 0. This well-known sign of multiple scattering \cite{Bee88} gets much stronger with increasing sample layer thickness.

The width of the broad Lorentzian is shown in \fref{fig:ms_SQw}d. An overall trend to larger widths with increasing $Q$ is visible on top of which a thicker sample size increases the width rather independently of $Q$. As can be seen in \fref{fig:ms_SQw}a, the data are at large energy transfers even flatter than the fit. This means that multiple scattering causes a very broad, background-like contribution.

\subsubsection{Evaluation of the intermediate scattering function with a stretched exponential.}

\begin{figure*}
\includegraphics[height=0.2\textwidth,width=0.25\textwidth,keepaspectratio=true]{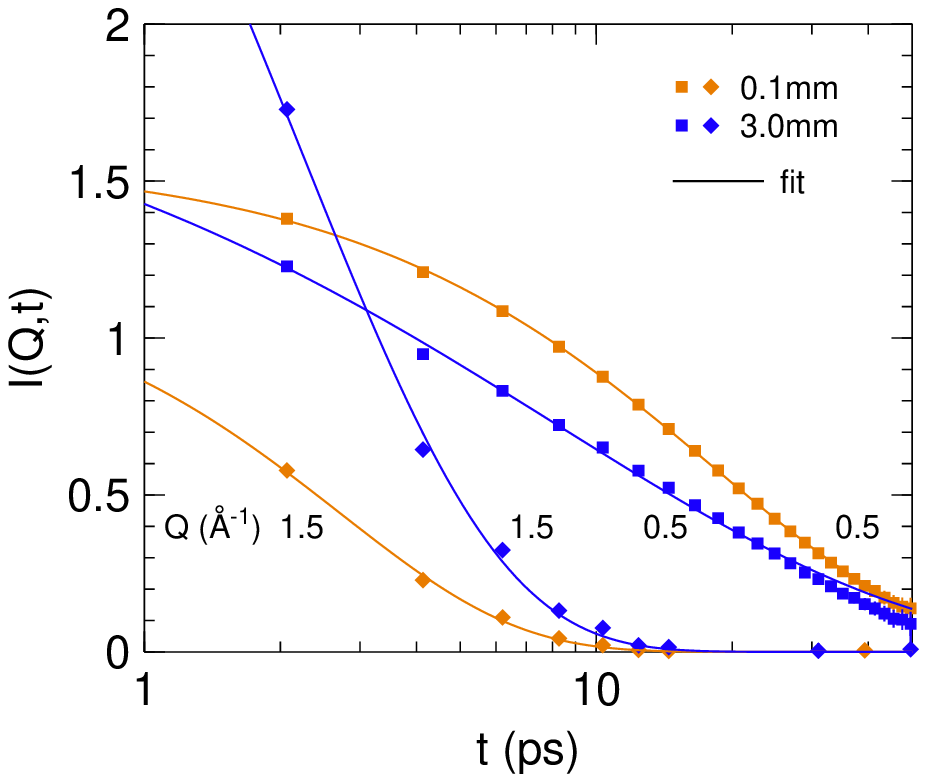}%
\hfill%
\includegraphics[height=0.2\textwidth,width=0.25\textwidth,keepaspectratio=true]{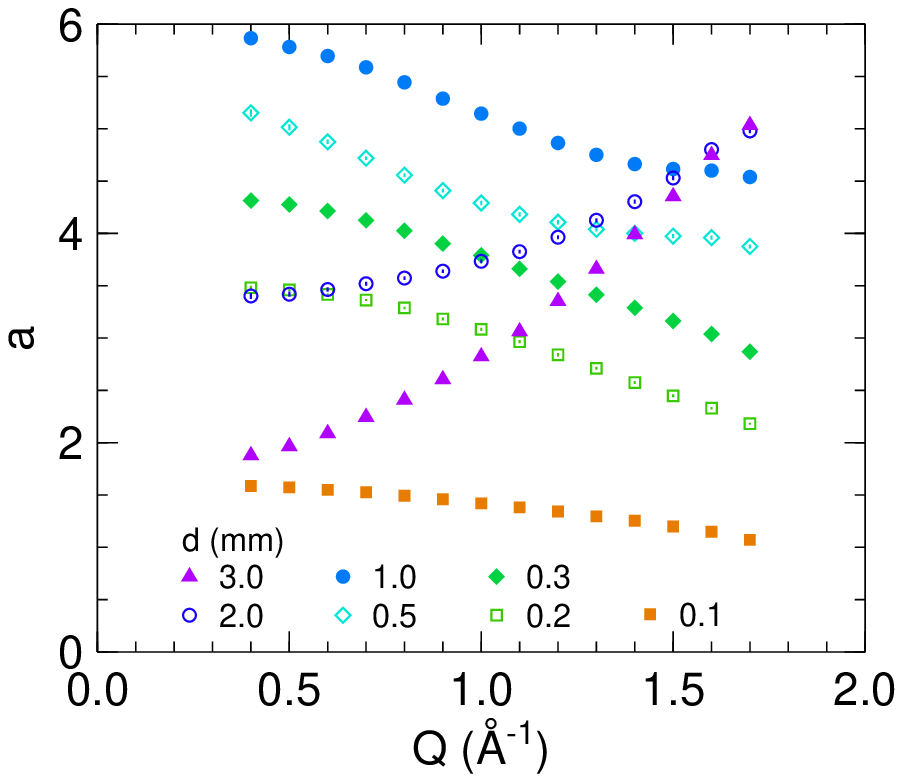}%
\hfill%
\includegraphics[height=0.2\textwidth,width=0.25\textwidth,keepaspectratio=true]{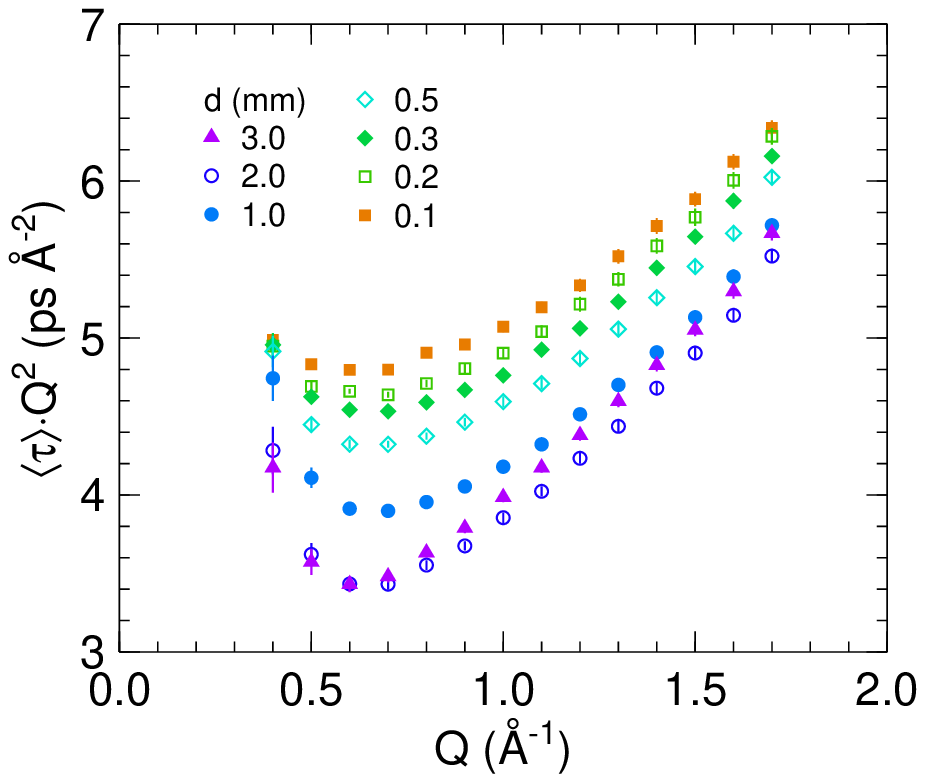}%
\hfill%
\includegraphics[height=0.2\textwidth,width=0.25\textwidth,keepaspectratio=true]{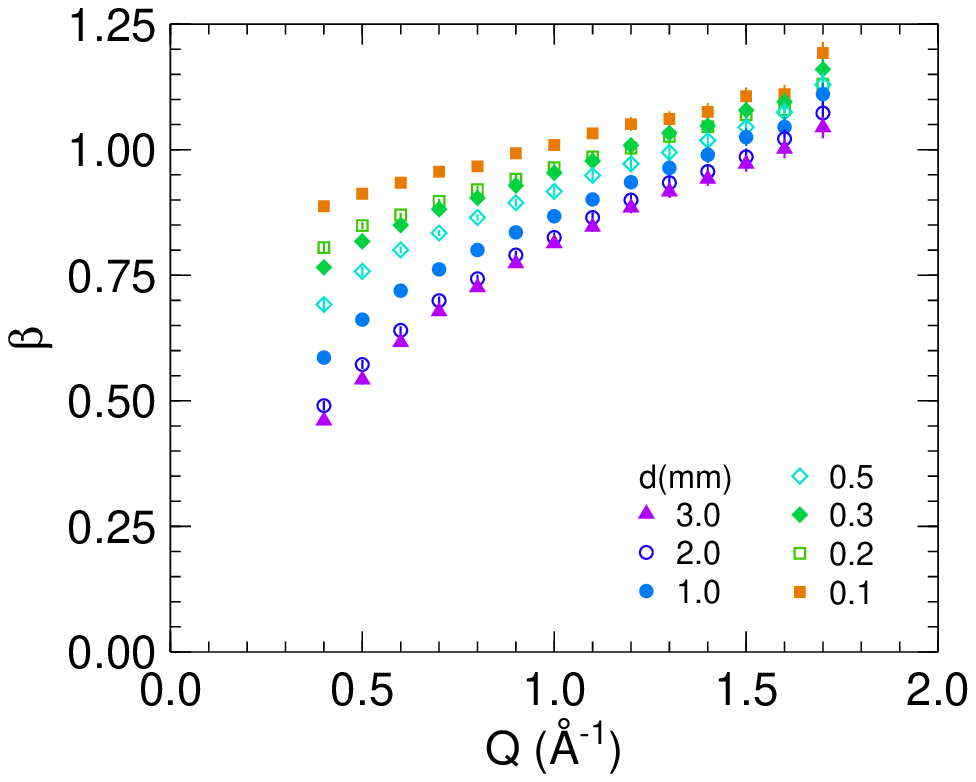}
\caption{From left to right: (a)~Intermediate scattering function of H$_2$O, shown at two $Q$ values and two sample layer thicknesses. Also shown is the fit of stretched exponentials to the data (drawn-out lines). (b)~The prefactor $a$. (c)~The mean relaxation time multiplied by $Q^2$. (d)~The stretching exponent $\beta$.}
\label{fig:ms_IQt}
\end{figure*}

The intermediate scattering function can be fitted well with a stretched exponential as can be seen in \fref{fig:ms_IQt}a.

The prefactor, displayed in \fref{fig:ms_IQt}b, is basically the value $I_\mathrm{inc}^\mathrm{exp}(Q,t=0)$, i.\,e.\ the zeroth Fourier component, which is the area under the scattering function, $\int_{-1\,\mathrm{meV}}^{1\,\mathrm{meV}} S_\mathrm{inc}^\mathrm{exp}(Q,\omega) \ \rmd \omega$. The limits of the integration are given by the region where the scattering function was evaluated -- which in turn is limited by the dynamical range of the experiment.

The mean relaxation time is influenced by multiple scattering, both in its average value and its $Q$ dependence. The strongest effect can be observed in \fref{fig:ms_IQt}c at intermediate $Q$ (around 0.7\,\AA$^{-1}$) where the apparent speed-up is most pronounced, in agreement with Monte Carlo simulations~\cite{Wuttke00mss}.

The stretching factor is more influenced than the mean relaxation time~\cite{Wuttke96}. Stretching increases ($\beta$ decreases) systematically, most at small values of $Q$, cf.\ \fref{fig:ms_IQt}d.

\subsection{Fitting artifacts}

\begin{figure*}
\includegraphics[height=0.375\textwidth,width=0.5\textwidth,keepaspectratio=true]{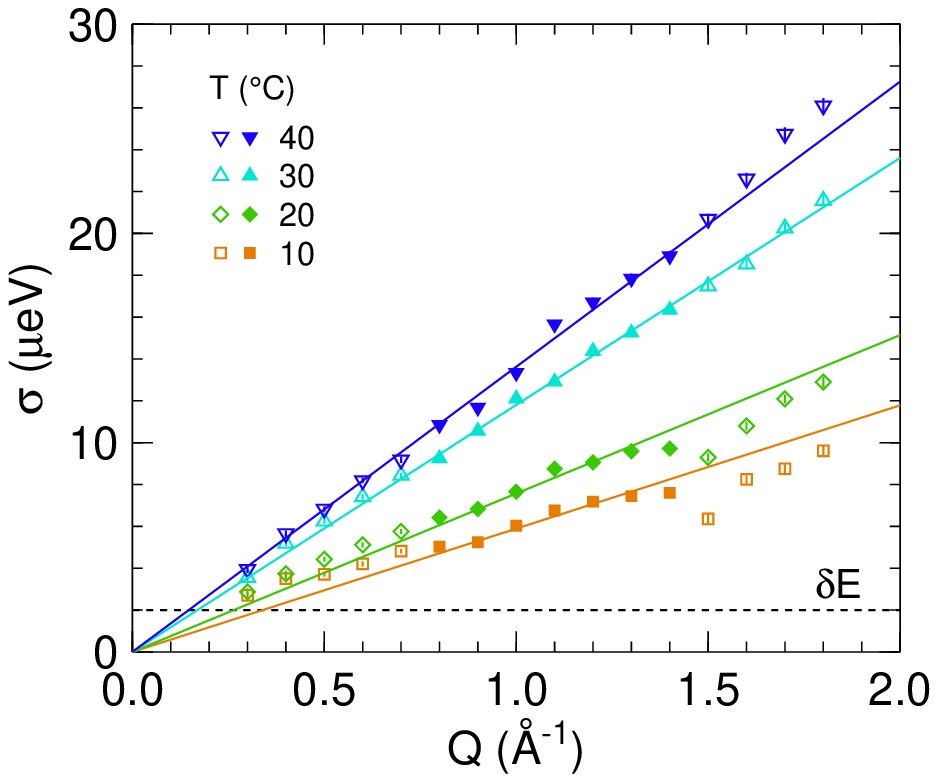}%
\hfill%
\includegraphics[height=0.375\textwidth,width=0.5\textwidth,keepaspectratio=true]{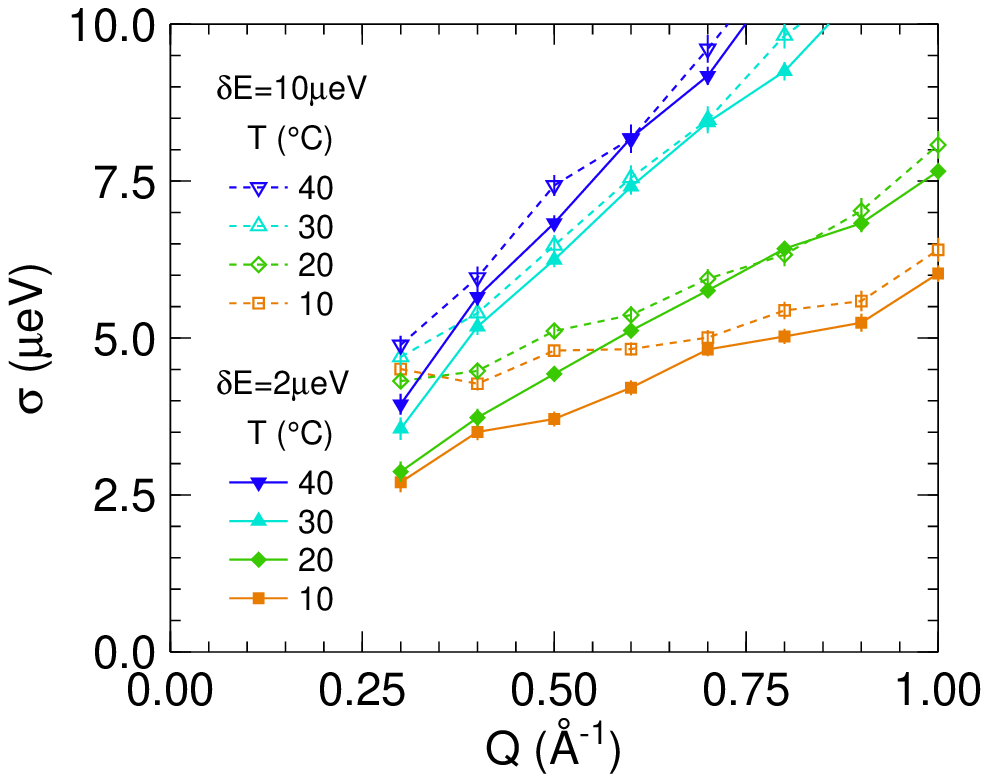}
\caption{From left to right: (a)~Width of the narrow Gaussian line of a three-component fit to the spectra of the fully hydrated phospholipid dimyristoylphosphatidylcholine (DMPC). The levelling off of the widths at low temperatures and small $Q$ (cf.\ \fref{fig:fit_w1}) is practically inexistent. The distance between two data points, $\delta E = 2$\,\textmu{}eV, is indicated in the figure. The energy resolution is about 60\,\textmu{}eV. The solid lines are fits to the filled points in the medium-$Q$ region. (b)~Close-up view of the low-$Q$ region of the width of the narrow Gaussian line of a three-component fit to the spectra of DMPC, measured with different point distances. It can be clearly seen that the effect of levelling off at small $Q$ can be nearly completely removed by choosing a higher point density.}
\label{fig:fit_w4}
\end{figure*}

The width of the narrow component of the fit to the phospholipid data with a binning of $\delta E = 10$\,\textmu{}eV is shown in \fref{fig:fit_w1}. The analogous fit to the data with $\delta E = 2$\,\textmu{}eV yields the line widths in \fref{fig:fit_w4}a. Also shown are in both figures fits of straight lines to the medium $Q$ range data points and the width of the bins.

It can be seen that in the case of the coarse energy binning shown in \fref{fig:fit_w1}, the data points deviate strongly from the linear behaviour at low $Q$ and level off to a value of about $\delta E / 2$. With a finer energy binning, \fref{fig:fit_w4}a, they display a much more linear behaviour.

This is shown in detail in \fref{fig:fit_w4}b where the fit results for the two energy binnings are compared. The effect is striking: the levelling off which was discussed to be either confinement or a fit artifact~\cite{Busch10flow} is clearly the latter.

\section{Discussion}

\subsection{Evaluation of the long-range motion}

Both, geometry and rate of the long-range motion are surprisingly little affected by multiple scattering. The rate of the diffusion is connected to the overall value of $\Gamma_\mathrm{narrow}/Q^2$ or $\langle \tau \rangle \cdot Q^2$, the geometry to the $Q$-dependence of these quantities.

The evaluation of the scattering function $S(Q,\omega)$ with the sum of two Lorentzians where the narrow accounts for the long-range motion yields very robust parameters for the narrow component.
As can be seen in \fref{fig:ms_SQw}b,
only for small $Q$, an increasing scattering power of the sample correlates with an increase of the line width.

Of the large deviation of the measured line width from the expected values for low $Q$ shown in \fref{fig:fit_w1}, up to 125\%, only a very small part can be attributed to multiple scattering: Extrapolating the 72\% and 85\% transmission measurements to one with 100\% transmission, we expect a deviation at small $Q$ of about 3\%. This extrapolation is known to be not suitable for the precise evaluation of multiple scattering~\cite{Slaggie67,Slaggie69} but it shows that the effects caused by multiple scattering are more than one order of magnitude lower than the observed deviations in \fref{fig:fit_w1}.

Much in contrast, the fitting artifacts distort the observed geometry of the long-range motion heavily. It will be shown in the following that a systematically too large line broadening (or too small correlation time) is obtained.

\begin{figure*}
\includegraphics[height=0.375\textwidth,width=0.5\textwidth,keepaspectratio=true]{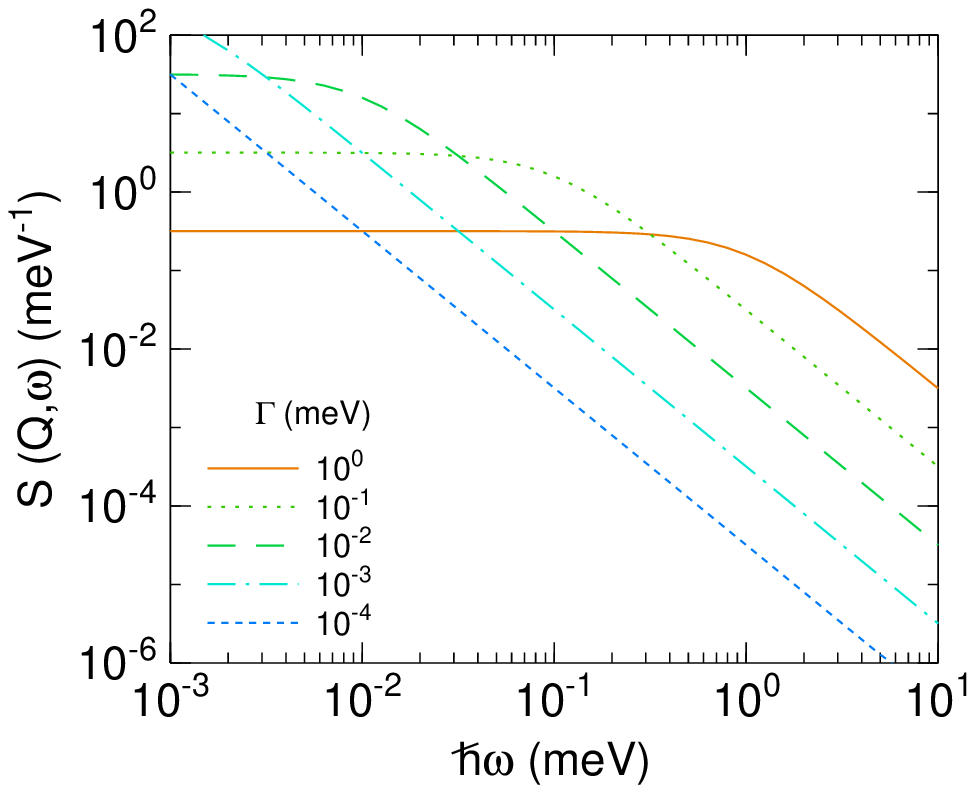}%
\hfill%
\includegraphics[height=0.375\textwidth,width=0.5\textwidth,keepaspectratio=true]{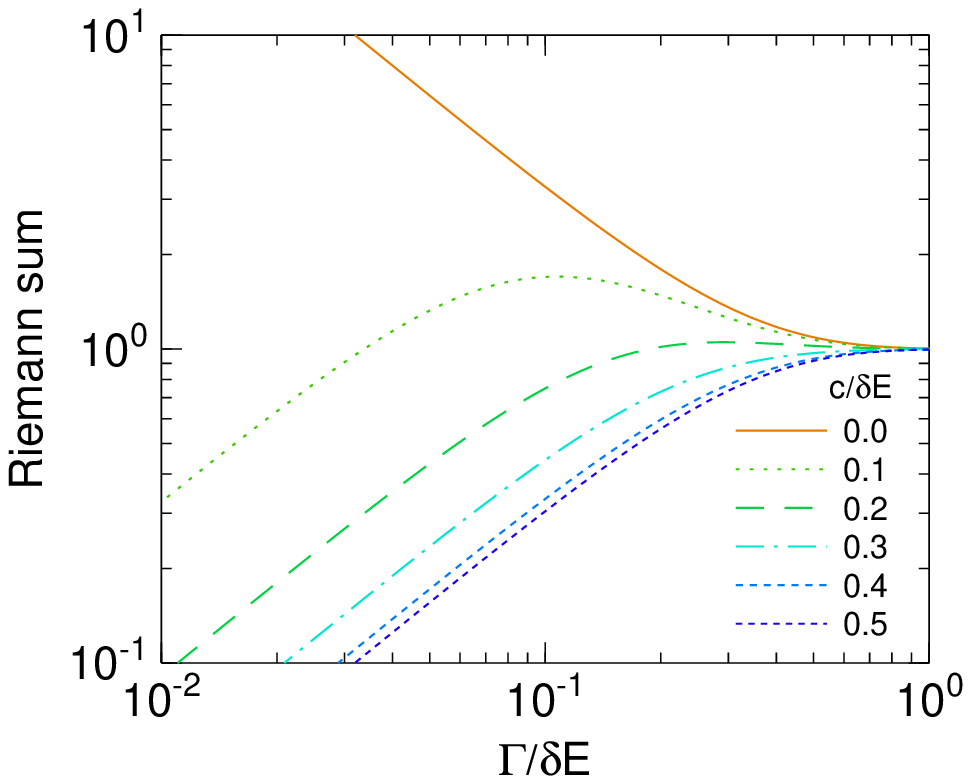}
\caption{From left to right: (a)~Lorentzian curves with different widths plotted double-logarithmically. After a plateau, the curves become a power law for $\hbar \omega \gtrsim \Gamma$. (b)~The Riemann sum over a Lorentzian as a function of its width $\Gamma$, shown for different values of their centers $c$. Both, width and center are given in units of the distance between two points at which the Lorentzians are evaluated $\delta E$. For $\Gamma \lesssim \delta E / 2$, the Riemann sum deviates significantly from the correct value of 1. The direction of the deviation depends on the relative position of the center of the Lorentzian with respect to the points.}
\label{fig:fit_art}
\end{figure*}

A Lorentzian line is practically indistinguishable from a power law for any $\hbar \omega \gtrsim \Gamma$, namely $(\Gamma/\omega)^2$, as shown in \fref{fig:fit_art}. This means that for any energy transfer bigger than $\Gamma$, the fit can be shifted to bigger $y$-values either by increasing the always present prefactor $a$ or -- completely equivalently -- by increasing $\Gamma$.

Increasing $a$ or increasing $\Gamma$ makes a difference only at energy transfers smaller than $\Gamma$: increasing $a$ increases the $y$-values in this range whereas increasing $\Gamma$ decreases them. It is therefore this range which determines $a$ and $\Gamma$.

The discrete convolution of the theoretical fit function $S_\mathrm{inc}^\mathrm{theo}(Q,\omega)$ with the instrumental resolution $R(Q,\omega)$ is the origin of this problem: whereas $R$ is in the present case very well-behaved, $S_\mathrm{inc}^\mathrm{theo}$ becomes so sharp peaked as $\Gamma$ decreases that the evaluated values at the points near zero energy transfer are not representative for the whole bin any more. This is because the curvature in the center of the Lorentzian is $-2/(\pi \Gamma^3)$ and diverges therefore for $\Gamma \rightarrow 0$.

The consequence is that the integral convolution in~\eref{equ:convint} cannot be substituted by the discrete convolution~\eref{equ:convsum} any more:
Whereas the integral under the Lorentzian is $\Gamma$-independently unity, the Riemann sum deviates very strongly when the line width is smaller than about half of the distance between two evaluated points, $\Gamma \lesssim \delta E / 2$.

It will be shown in the following that the systematically too large fit value for the line width does not depend on the direction of this deviation in the critical cases discussed here ($\Gamma \approx 1$\,\textmu{}eV, $\delta E = 10$\,\textmu{}eV):

(a)~If the sum is calculated too big (the center of the Lorentzian is close to a measured point), the central points are mistakenly calculated too high. Only the central point which is evaluated at the center of the Lorentzian is in the plateau region of the Lorentzian.

(a.1)~The fit can increase $\Gamma$ and hereby lower the central points by both, the natural decrease (\fref{fig:fit_art}a) and the better calculation (\fref{fig:fit_art}b). Simultaneously, the function values in the wings of the Lorentzian will increase (\fref{fig:fit_art}a) which will be compensated by a smaller value for the prefactor $a$, making only a small counter-contribution to lower the central points.

(a.2)~The fit could alternatively decrease the prefactor to get the central points down. This leads to a fit curve that is systematically below the data points at large energy transfers -- it will be shifted upwards by increasing $\Gamma$.

(b)~If the sum is calculated too small (the center of the Lorentzian is close to the middle in between two measured points), the central points are mistakenly calculated too low. The points which are evaluated close to the center of the Lorentzian are already in the power law wing region of the Lorentzian.

(b.1)~The fit can increase $\Gamma$ and hereby increase the central points by both, the natural increase (\fref{fig:fit_art}a) and the better calculation (\fref{fig:fit_art}b). Simultaneously, the function values in the wings of the Lorentzian will increase (\fref{fig:fit_art}a) which will be compensated by a smaller value for the prefactor $a$ which cannot compensate the increase caused by the better calculation though.

(b.2)~The fit could alternatively increase the prefactor to get the central points up. This leads to a fit curve that is systematically above the data points at large energy transfers -- it would need to be shifted downwards by decreasing $\Gamma$. However, this would yield even lower central points (\fref{fig:fit_art}b) and is therefore not an option for the fit.

In conclusion, the fit will \emph{always} yield a line width $\Gamma$ that is too broad if $\Gamma \gtrsim \delta E /2$. In the example of the phospholipid measurements shown in \fref{fig:fit_w1} and \fref{fig:fit_w4}, the deviation from a linear behaviour at small $Q$ is decreased from 125\% (in \fref{fig:fit_w1}) to about 30\% (in \fref{fig:fit_w4}) by choosing a smaller energy step between measured points.

This 30\% deviation might very well be due to the background in the resolution measurement, caused by the instrument and phonons in the vanadium. Comparing the fits using a convolution with the instrumental resolution as determined by a vanadium measurement at temperatures between 4\,K and 300\,K as well as Gaussians with flat background, the fitted values differed in this order of magnitude without showing clear trends. This will be studied in more detail and presented in a forthcoming publication.
\subsection{Evaluation of the localized motions}

The evaluation of the localized motions in the two-component fit of the scattering function is influenced in both, rate and geometry. This is because the multiply scattered neutrons obtain relatively large energy transfers, yielding a nearly flat background in the energy direction. With an increasing amount of multiple scattering, this background increases. The broad Lorentzian which accounts for localized motions therefore broadens and gets more intensity.

The geometry is increasingly hard to evaluate from the EISF as it does not approach 1 for small momentum transfers any more.
It seems to be the most sensitive indicator for multiple scattering.
Including an arbitrary scaling factor in the definition of the EISF as done previously~\cite{Busch10flow} makes an approximate evaluation possible at reasonable transmissions~\cite{Wuttke00mss}.

In the following, we will try to explain the effects of multiple scattering with simple phase space arguments, assuming for the calculation of $Q$ that the neutron does not change its energy in a scattering process and neglecting the geometry of the sample~\cite{Busch07dipl}. Knowing that the spectrum of multiply scattered neutrons is essentially a convolution of the double differential cross section with itself for every scattering process~\cite{Copley74} and that two times scattering is the most probable multiple scattering event~\cite{Settles96phd}, the $Q$ values of the two scattering events are the interesting quantities. The width of the observed scattering function will then be approximatively the sum of the two single scattering processes.

\begin{figure*}
\includegraphics[height=0.375\textwidth,width=0.5\textwidth,keepaspectratio=true]{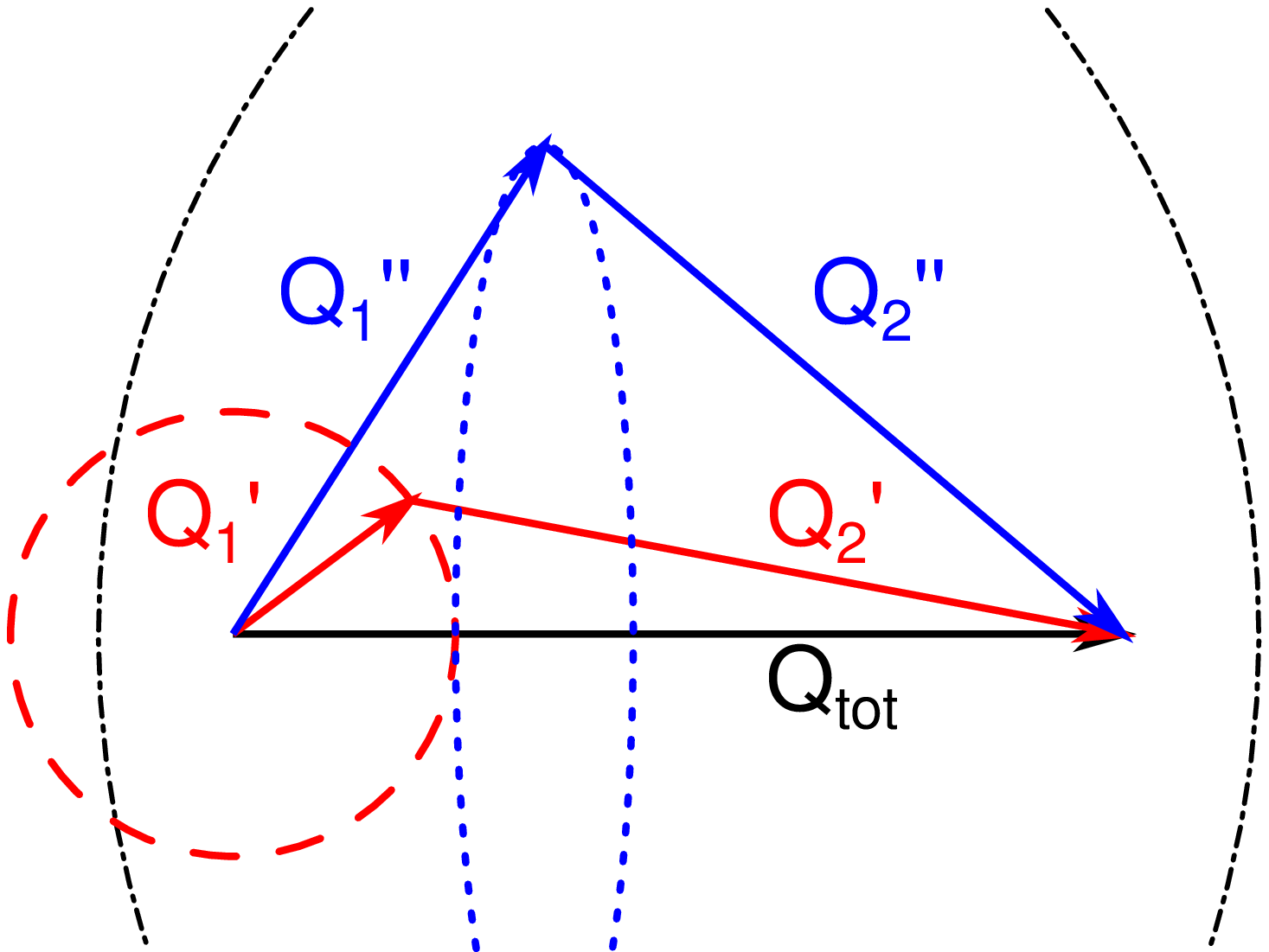}%
\hfill%
\includegraphics[height=0.375\textwidth,width=0.5\textwidth,keepaspectratio=true]{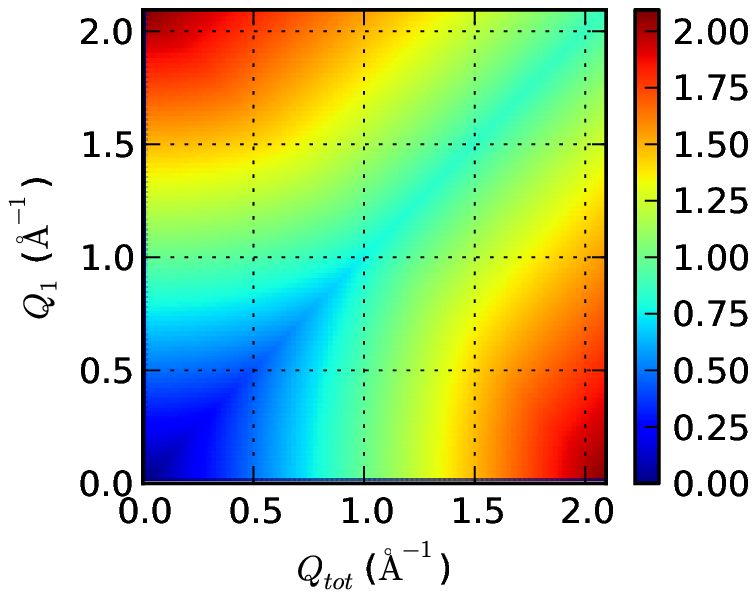}
\caption{From left to right: (a)~Two scattering processes, denoted with subscript 1 and 2, with their respective scattering vectors $\vec{Q}_1$ and $\vec{Q}_2$ combine to a total scattering process with $\vec{Q}_\mathrm{tot}$. The black dash-dotted lines denote the external borders in between which a connection point between the two scattering events must be so that the given $\vec{Q}_\mathrm{tot}$ can be reached with two quasi-elastic scattering processes. Red dashed line: section of the sphere on which $\vec{Q}_{1}'$ may end and the paper plane. Note that only points on this sphere which are within the black dash-dotted line can contribute to the shown scattering process. Blue dotted line: projection of the circle (which is perpendicular to the paper plane) on which the connection point between the two $\vec{Q}''$ can revolve. (b)~Mean $Q_2$ for a given combination of $Q_1$ and $Q_\mathrm{tot}$.}
\label{fig:ms_q_comb}
\end{figure*}

Using Monte Carlo simulations~\cite{Wuttke00mss}, it was determined that the
probability distribution of the scattering angles of the single scattering processes is -- independently of the total scattering angle -- nearly triangular with a maximum probability of 90\textdegree{}, corresponding to a $Q$ of already 0.71$\cdot Q_\mathrm{max}$ (elastic), in this case to 1.5\,\AA$^{-1}$. A similar assessment has also been used elsewhere~\cite{Cusack90}.
Half of the neutrons will be scattered with a larger $Q$
and result in a very broad line contributed by the multiply scattered neutrons.

The upper limit for the values of $Q$ that can possibly participate in a double scattering event is given by the wavelength of the neutrons: $Q_{1,2}^\mathrm{max} = 4 \pi / \lambda$. The lower limit for $Q_{1,2}$ seen from a given $Q_\mathrm{tot}$ can be deduced from \fref{fig:ms_q_comb}a where the vectorial momentum transfer of the first scattering process is denoted with $\vec{Q}_1$, the one of the second with $\vec{Q}_2$, and the total (observed) by $\vec{Q}_\mathrm{tot} = \vec{Q}_1 + \vec{Q}_2$. This means for the absolute values: $Q_1 + Q_2 \geqslant Q_\mathrm{tot}$.
The distribution of possible $Q$ values in the scattering process and therefore of relaxation times is therefore largest at small $Q_\mathrm{tot}$ (cf.\ \fref{fig:ms_IQt}d).

The shift with regards to the scattering angle and the cut-off of $Q$ by the choice of the wavelength of the incoming neutrons brings about that a change of this wavelength influences multiple scattering not only for coherently scattering samples~\cite{Sears75,Russina99,Mezei99,Wuttke00mss} but also for incoherently scattering samples~\cite{Settles96phd,Wuttke00mss}, although to a smaller extent.

To understand the resulting line width,
one needs the conditional probability for the two scattering events given some $Q_\mathrm{tot}$. We have calculated the mean $Q_2$ after a given $Q_1$ as follows (cf.\ \fref{fig:ms_q_comb}a): The probability for a certain ($Q_1, Q_2$)-combination is directly proportional to the number of allowed ($\vec{Q}_1, \vec{Q}_2$)-combinations, weighed with the corresponding solid angles.

The probability to have an event with a certain $\vec{Q}_{1,2}$ is inversely proportional to the surface of a sphere with radius $Q_{1,2}$, $1/(4 \pi Q_{1,2}^2)$.
Multiplying these two probabilities gives the one to reach a certain $\vec{Q}_\mathrm{tot}$ via a given connection point of $\vec{Q}_1$ and $\vec{Q}_2$.
Each of the two scattering events have to be multiplied with the solid angle accessible with a given scattering angle, $\sin(2\theta_{1,2})/\lambda$.

As all connection points which satisfy $Q_{1,2} \leq Q_\mathrm{max}$ can be used equivalently, this probability has to be multiplied with the circumference $l$ of the circle around $\vec{Q}_\mathrm{tot}$,
\begin{eqnarray}
    \fl
    l = \frac{\pi}{Q_\mathrm{tot}} \cdot \Re \Big\{ \big[ -(Q_\mathrm{tot}-Q_1-Q_2) \cdot \nonumber \\
    \fl
    \qquad\qquad
    (Q_\mathrm{tot}-Q_1+Q_2) \cdot \nonumber \\
    \fl
    \qquad\qquad
    (Q_\mathrm{tot}+Q_1-Q_2) \cdot \nonumber \\
    \fl
    \qquad\qquad
    (Q_\mathrm{tot}+Q_1+Q_2) \big]^{1/2} \Big\} \quad ,
    \label{equ:cut_spheres}
\end{eqnarray}
where $\Re$ denotes the real part of the argument.

The result, symmetrical in $Q_1$ and $Q_2$, is displayed in \fref{fig:ms_q_comb}b. It can be seen that even if one scattering process has already $Q_1 = Q_\mathrm{tot}$, the other scattering process still has a nonzero $Q_2$. Two limiting behaviours can be observed: On the one hand, for $Q_1$ and $Q_\mathrm{tot} \gtrsim 1.0$\,\AA$^{-1}$, the mean $Q_2$ depends basically only on $|Q_\mathrm{tot}-Q_1|$ and is never less than about 1.0\,\AA$^{-1}$. On the other hand, for $Q_1$ and $Q_\mathrm{tot} \lesssim 1.0$\,\AA$^{-1}$, we observe $Q_2 \approx \max(Q_1, Q_\mathrm{tot})$. Scattering events where the second scattering process has a very small $Q$ compared to $Q_1$ and $Q_\mathrm{tot}$, i.\,e.\ broadening the signal only slightly, are very rare. This causes the distinctness between the signal and the much broader contribution of multiply scattered neutrons.

\subsection{Phospholipid motions}

It is clear that the long-range motion extracted from a measurement with an observation time of 60\,ps is not the macroscopic long-range motion. This manifests for example in a higher velocity~\cite{Busch10flow} and lower influence of additives~\cite{Busch11add} than observed macroscopically. Although the molecule travelled only a distance of about 1\,\AA{}~\cite{Busch10flow} during this short time, the interactions with the neighbours, the cage, are already established. The observed flow-like motion is not the initial phase of free flight before the molecule experiences the presence of the surrounding neighbours. This is evidenced by both, the clear existence of localized (caged) motions and the small flow velocity which is nearly three orders of magnitude below the thermal expectation value~\cite{Busch10flow}.

As shown in the present contribution, the neutron scattering data show no significant evidence for a restriction of the flow-like long-range motion to a cage within the observation time -- above and below the main phase transition.

We explain this with ``floppy modes'': the molecules are pushed around by density fluctuations resembling longitudinal acoustic phonons in crystals. Below the phase transition, a restoring force will eventually push them back to the original place. Due to the ``floppiness'' of the material, the molecules travel more than the 1\,\AA{} during which the experiment follows them before the restoring force gets big enough to restrict their motion. Therefore, this effect does not leave a trace in the data. Above the phase transition, the molecules are transported by many short subsequent uncorrelated flow events.

These low-frequency ``floppy modes'' are known from disordered colloidal crystals~\cite{Kaya10}, colloidal glass formers, and supercooled liquids and signal the onset of macroscopic elasticity~\cite{Ghosh10scl}. Phospholipid membranes also exhibit elasticity above and below the main phase transition~\cite{Harland10} and the observed flow patterns~\cite{Falck08} resemble very much dynamical heterogeneities~\cite{Busch11add}, a hallmark of glassy dynamics, which were again suggested to be linked with the ``floppy modes''~\cite{Donati98}.

The reason that the dynamics of phospholipids resembles so much dense, supercooled systems is probably the very high density in the head group area of the phospholipids~\cite{Busch11add}. This similarity has proven to be extremely useful already in the past when the free volume theory developed for glasses~\cite{Cohen59} was adapted to membranes~\cite{Galla79walk,Vaz91}. Transposing the newer concepts developed in glass physics, dynamical heterogeneities and ``floppy modes'', to phospholipid membranes seems not only consequent but can also explain the observed features effortless.

\section{Summary and conclusions}

The molecular motion of the phospholipid DMPC was studied below and above the main phase transition on a time scale of about 60\,ps, taking the influence of multiple scattering and fit artifacts into particular consideration.

Multiple scattering was shown to be only a minor nuisance for the determination of the long-range motion of an incoherent scatterer. Exemplary, H$_2$O samples with transmissions between 85\% and as low as 2\% were studied. The multiply scattered neutrons are lost in the elastic region of the signal and create a background which is a very broad line in energy space. In a multicomponent analysis, they affect therefore only the evaluation of localized motions. This, together with the excellent agreement of the results with Monte Carlo simulations shows that one could more often dare to use strongly scattering samples for quasielastic measurements.

The fit of a theoretical scattering function to measured data after numerical convolution with a measured resolution function can give erroneous results. A delicate combination of correlated fitting parameters and a discretization problem during the convolution with the measured instrumental resolution causes an error in an unexpected place: the line width. Analyzing the data with different bin widths should give indications if this problem is present. The lower limit for the bin width is often given by the online histogramming of the neutrons during the experiment -- a complete event recording of the detected neutrons, in principle possible, would remove this limitation.

Flow-like motions of the phospholipid molecules bear a striking resemblance to dynamical heterogeneities above the main phase transition and to ``floppy modes'' below. The confinement of the molecules below the main phase transition is not visible in the long-range component of the fit due to the short observation time of the experiment. It seems that the description of motions in the phospholipid membrane could much profit from transferring the up-to-date concepts used in glass physics.

\ack{We would like to express our thankfulness to Prof.~W.~Petry who has extensively supported our work. We enjoyed the lively discussions with him and his special interest in the here-discussed multiple scattering effects. We have very much profited from the earlier work of and many discussions with J.~Wuttke who first pointed out the importance of the numerical convolution to us. We would like to thank A.~Stadler for discussions; G.~Brandl, F.~Lipfert, B.~Mihiretie, and A.~Tischendorf for their participation in the measurements of H$_2$O; and H.~Morhenn for performing the phospholipid measurements.}

\end{document}